\documentclass[conference]{IEEEtran}

\usepackage{hyperref}

\ifCLASSINFOpdf
\else
\fi

\usepackage{amsmath}
\usepackage{verbatim}
\usepackage{fancyhdr}
\usepackage{array}
\usepackage{amsmath}
\usepackage{amssymb}
\usepackage{amsthm}
\usepackage{mathrsfs,bbm}
\usepackage{eqnarray}
\usepackage{cancel}
\usepackage{multirow}

\def \E {\mathbf{E}}
\def \as {_\mathrm{a.s.}}

\newcommand{\RR}{\mathbb{R}}
\newcommand{\NN}{\mathbb{N}}

\def \ud {\,\mathrm{d}}

\DeclareMathOperator*{\argmax}{argmax}

\theoremstyle{plain}
\newtheorem{theorem}{Theorem}
\newtheorem*{theorem*}{Theorem}
\newtheorem*{conjecture*}{Conjecture}
\newtheorem{lemma}{Lemma}

\newtheorem*{claim*}{Claim}
\newtheorem*{corollary*}{Corollary}

\theoremstyle{definition}
\newtheorem{definition}{Definition}

\theoremstyle{remark}

\newtheorem*{remarks*}{Remarks}

\begin{document}
%
\title{Canonical Estimation in a Rare-Events Regime}

\author{\IEEEauthorblockN{Mesrob I. Ohannessian}
\IEEEauthorblockA{Laboratory for Information\\ and Decision Systems\\
Massachusetts Institute of Technology\\
Cambridge, MA 02139\\
Email: \href{mailto:mesrob@mit.edu}{mesrob@mit.edu}}
\and
\IEEEauthorblockN{Vincent Y. F. Tan}
\IEEEauthorblockA{Department of \\ Electrical and Computer Engineering\\
University of Wisconsin-Madison\\
Madison, WI 53706\\
Email: \href{mailto:vtan@wisc.edu}{vtan@wisc.edu}}
\and
\IEEEauthorblockN{Munther A. Dahleh}
\IEEEauthorblockA{Laboratory for Information\\ and Decision Systems\\
Massachusetts Institute of Technology\\
Cambridge, MA 02139\\
Email:\href{mailto:dahleh@mit.edu}{\ dahleh@mit.edu}}}


%


\maketitle

\begin{abstract}
We propose a general methodology for performing statistical inference within a `rare-events regime' that was recently suggested by Wagner, Viswanath and Kulkarni. Our approach allows one to easily establish consistent estimators for a very large class of canonical estimation problems, in a large alphabet setting. These include the problems studied in the original paper, such as entropy and probability estimation, in addition to many other interesting ones. We particularly illustrate this approach by consistently estimating the size of the alphabet and the range of the probabilities. We start by proposing an abstract methodology based on constructing a probability measure with the desired asymptotic properties. We then demonstrate two concrete constructions by casting the Good-Turing estimator as a pseudo-empirical measure, and by using the theory of mixture model estimation.
\end{abstract}


%
\IEEEpeerreviewmaketitle

\section{Introduction}

We propose a general methodology for performing statistical inference within the `rare-events regime' suggested by Wagner, Viswanath and Kulkarni in \cite{wext:wvk}, referred to as WVK hereafter. This regime is a scaling statistical model that strives to capture large alphabet settings, and is  characterized by the following notion of a \emph{rare-events source}.

\begin{definition} \label{def:rare-events}
Let $\{(A_n,p_n)\}_{n\in\NN}$ be a sequence of pairs where each $A_n$ is an alphabet of finite symbols, and $p_n$ is a probability mass function over $A_n$. Let $X_n$ be a single sample from $p_n$, and use it to define a `shadow' sequence $Z_n=n p_n(X_n)$. Let $P_n$ denote the distribution of $Z_n$. We call  $\{(A_n,p_n)\}_{n\in\NN}$ a \emph{rare-events source}, if the following conditions hold.
\begin{itemize}
\item[(i)] There exists an interval $C=[\check{c},\hat{c}]$, $0<\check{c}\leq\hat{c}<\infty$, such that for all $n\in\NN$ we have $\frac{\check c}{n}\leq p_n(a) \leq\frac{\hat c}{n}$ for all $a\in A_n$, or equivalently, $P_n$ is supported on $C$.
\item[(ii)] There exists a random variable $Z$, such that $Z_n\to Z$ in distribution. Equivalently, there exists a distribution $P$, such that $P_n \Rightarrow P$ weakly.
\end{itemize}
\end{definition}

To complete the model, we adopt the following sampling scheme. For each $n$, we draw $n$ independent samples from $p_n$, and we denote them by $X_{n,1},\cdots,X_{n,n}$. Using these samples, we are interested in estimating various quantities. WVK consider, among a few others, the following:
\begin{itemize}
\item The total (Good-Turing) probabilities of all symbols appearing exactly $k$ times, for each $k\in\NN_0$.
\item The normalized log-probability of the observed sequence.
\item The normalized entropy of the source.
\item The relative entropy between the true and empirical distributions.
\end{itemize}
They also consider two-sequence problems and hypothesis testing, but we focus here on single sequence estimation.

It is striking that many of these quantities can be estimated in such a harsh scaling model, where one cannot hope for the empirical distribution to converge in any traditional sense. However, WVK's estimators have some drawbacks. For example, since they are based on series expansions of the quantities to be estimated, one has to carefully choose the growth rate of partial sums, in order to control convergence properties. More importantly, they are specifically tailored to each individual task. Their consistency is established on a case-by-case basis. What is desirable, and what this paper contributes to, is a methodology for performing more general statistical inference within this regime. Ideally such a framework would allow one to tackle a very large class of canonical estimation problems, and establish consistency more easily. 

We may summarize the fundamental ideas behind our approach and the organization of this paper as follows. First, in Section \ref{sec:problems}, we isolate the class of estimation problems that we are interested in as those that asymptotically converge to an integral against $P$. The quantities studied by WVK fall in this category, and so do other interesting problems such as estimating the size of the alphabet. Other problems, such as estimating the range of the probabilities given by the support interval $C$, can also be studied in this framework.

Next, in Section \ref{sec:solution}, we propose an abstract solution methodology. At its core, we construct a (random) distribution $\tilde P_n$ that converges weakly to $P$ for almost every observation sample. This construction immediately establishes the consistency of natural estimators for the abovementioned quantities, if bounds on $C$ are known. If in addition the rate of the convergence of $\tilde P_n$ is established, the framework gives consistent estimators even without bounds on $C$.

To make this methodology concrete, we build on a core result of WVK that establishes the strong consistency of the Good-Turing estimator. In particular, since the role of the empirical measure is lost, we show in Section \ref{sec:pseudo-empirical} that we can treat the Good-Turing estimator as a pseudo-empirical measure. Once this is established, we can borrow heavily from the theory of mixture models, where inference is done using i.i.d.\ samples, and adapt it to our framework. In Section \ref{sec:weakly-convergent}, we suggest two approaches for constructing $\tilde P_n$: one that is based on maximum likelihood, and another that is based on minimum distance. Both constructions guarantee the almost sure weak convergence of $\tilde P_n$ to $P$, but the latter, under some conditions, also provides the desirable convergence rates.

In Section \ref{sec:applications} we illustrate the methodology with some examples. In particular, we show how one can consistently estimate the entropy of the source and the probability of the sequence as studied by WVK, but we also propose consistent estimators for the size of the alphabet and for the support interval $C$.

\subsubsection*{Notation}
Throughout, we use $F(.;.)$ to denote the cumulative distribution of the second argument (which is a probability measure on the real line or on the integers) evaluated at the first argument (which is a point on the real line or an integer).

\section{A General Class of Estimation Problems } \label{sec:problems}

\subsection{Definitions}

Given i.i.d.\ samples $X_{n,1},\cdots,X_{n,n}$ from the rare-events source $(A_n,p_n)$, we can pose a host of different estimation problems. Since the alphabet is changing, quantities that depend on explicit symbol labels are not meaningful. Therefore, one ought to only consider estimands that are invariant under re-labeling of the symbols in $A_n$. In particular, we consider the following class of general estimation problems.

\begin{definition} \label{def:canonical} Consider the problem of estimating a sequence $\{Y_n\}_{n\in \NN}$ of real-valued random variables  using, for every $n$, the samples $X_{n,1},\cdots,X_{n,n}$. We call this a \emph{canonical estimation problem} if, for every rare-events source, we have:
\begin{equation} \label{eq:canonical}
\E\left[Y_n\right] = \int_C f_n(x) \ud P_n(x).
\end{equation}
for some sequence $\{f_n\}$ of continuous real-valued functions on $\RR^+$ that converge pointwise to a continuous function $f$.
\end{definition}

It is worth noting that it follows that $\{f_n\}$ and $f$ are also bounded on every closed interval $[a,b]$, $0<a\leq b<\infty$. Observe that this definition corresponds indeed to estimands that are invariant under re-labeling, in expectation. The following lemma characterizes the limit. 
\begin{lemma} \label{lemma:limit}
For any canonical estimation problem,
\begin{equation}
	\E[Y_n] \to  \int_C f(x) \ud P(x).
\end{equation}
\end{lemma}
\begin{IEEEproof}
Since $P_n \Rightarrow P$, we can apply Skorokhod's theorem (\cite{wext:billingsley}, p. 333), to construct a convergent sequence of random variables $\xi_n\to\as\xi$, where $\xi_n  \sim P_n$ and $\xi \sim P$. By continuity, it follows that $f_n(\xi_n)\to\as f(\xi)$. By the bounded convergence theorem, we then have  $\E[f_n(\xi_n)]\to \E[f(\xi)]$. Since $\E[Y_n]=\E[f_n(\xi_n)]$, and  $\int_C f(x) \ud P(x)=\E[f(\xi)]$, the lemma follows.
\end{IEEEproof}

It is often more interesting to consider the subclass of canonical problems where there is strong concentration around the mean, and where the Borel-Cantelli lemma applies to give almost sure convergence to the mean. 

\begin{definition}
If a canonical estimation problem further satisfies $\left|Y_n-\E[Y_n]\right|\to\as 0$, then call it a \emph{strong canonical problem}. It follows that for strong canonical problems, 
\begin{equation}
	Y_n \to\as  \int_C f(x) \ud P(x).
\end{equation}
\end{definition}

Using these definitions, a reasonable estimator will at least agree with the limit set forth in Lemma \ref{lemma:limit}. Other modes of convergence may be reasonable, but we would like to exhibit a statistic that almost surely converges to that limit. We make this precise in the following definition.

\begin{definition} 
Given a canonical problem as in Definition \ref{def:canonical}, a corresponding \emph{estimator} is a sequence $\{\hat Y_n\}_{n\in \NN}$ such that, for each $n$, $\hat Y_n(a_{1},\cdots,a_{n})$ is a real-valued function on $\left(A_n\right)^n$, to be evaluated on the sample sequence $X_{n,1},\cdots,X_{n,n}$. A \emph{consistent estimator} is one that obeys
\begin{equation} \label{eq:performance}
	\hat Y_n(X_{n,1},\cdots,X_{n,n}) \to\as  \int_C f(x) \ud P(x).
\end{equation}
\end{definition}

For canonical estimation problems that are not necessarily strong, this approach produces an asymptotically unbiased estimator, with asymptotic mean squared error that is no more than the asymptotic variance of the estimand itself. For strong canonical estimation problems, this approach establishes strong consistency, in the sense that the estimator converges to the estimand, almost surely.

\subsection{Examples} \label{sec:examples}

To motivate the setting we have just described, we first note that all of the quantities studied by WVK are strong canonical estimation problems. For each quantity, WVK propose an estimator, and individually establish its consistency by showing almost sure convergence to the limit in Lemma \ref{lemma:limit}. In contrast, what we emphasize here is that this can potentially be done \emph{universally} over all strong canonical problems.

To highlight the usefulness of this generalization, we illustrate two important quantities that fall within this framework. We will revisit these in more detail in Section \ref{sec:applications}. The first quantity is the normalized size of the alphabet: $|A_n|/n$. For this, one can show (see, for example, \cite{wext:bhat}), that $|A_n|/n = \int_C \frac{1}{x} \ud P_n(x)$. Therefore we can take $f_n(x)=f(x)=\frac{1}{x}$, and since the estimand is deterministic, we have a strong canonical estimation problem.

The second quantity of interest is the interval $C$, or equivalently its endpoints $\check{c}$ and $\hat{c}$. Note that, by construction, $P$ is supported on $C$. Without loss of generality, we may assume that $\check{c}$ and $\hat{c}$ are respectively the essential infimum and essential supremum of $Z\sim P$. Therefore, note that $\left(\int x^{\pm q} \ud P(x)\right)^{1/q}$  converges to the essential infimum ($-$) or supremum ($+$) as $q\to\infty$. We can therefore consider, for fixed $q\ge 1$, the strong canonical problems that ensue from the choices $f_n(x)=f(x)=x^{-q}$ and $f_n(x)=f(x)=x^{q}$. These, by themselves, are not sufficient to provide estimates for $\check{c}$ and $\hat{c}$. However if, in addition to consistency, we establish the convergence rates of their estimators, then we can apply our framework to estimate $C$, as we show in Section \ref{sec:applications}.

\section{Solution Methodology} \label{sec:solution}

Our task now is to exhibit consistent estimators to canonical problems. We present here our abstract methodology, which we demonstrate concretely in Section \ref{sec:weakly-convergent}. The core of our approach consists of using the samples $X_{n,1},\cdots,X_{n,n}$ to construct a random measure $\tilde P_n$ over $\RR^+$, such that for almost every sample sequence, the sequence of measures $\{\tilde P_n\}$ converges weakly to $P$. We write: as $n\to\infty$
\vspace{-2pt}
\begin{equation} \label{eq:as-weak}
	\tilde P_n\Rightarrow\as P.
\end{equation}

If we accomplish this, we can immediately suggest a consistent estimator under certain conditions, as expressed by Lemma \ref{lemma:setting1}. We will be interested in integrating functions against the measure $\tilde{P}_n$. However, since the support $C$ of $P$ is unknown, we first introduce the notion of a \emph{tapered function} as a convenient way to control the region of integration. Given a real-valued function $g(x)$ on $\RR^+$, for every $D\geq 1$ define its $D$-tapered version as:
\begin{equation*}
g_D(x) \equiv \left\{
\begin{array}{lcl}
g(D^{-1})&\quad&x<D^{-1}\\
g(x)&\quad&x\in[D^{-1},D]\\
g(D)&\quad&x>D
\end{array}
\right.
\end{equation*}
If $g$ is continuous on $(0,+\infty)$, then we can think of $g_D(x)$ as a bounded continuous extension of the restriction of $g$ on $[D^{-1},D]$ to all of $\RR^+$. 

\begin{lemma} \label{lemma:setting1}
Consider a canonical problem characterized by some $f$. Let the support $C$ of a rare-events source be known up to an interval $[D^{-1},D]\supseteq C$ for some $D>1$. Then, if $\tilde P_n\Rightarrow\as P$ as $n\to\infty$, we have that
\begin{equation}
	\hat Y_n = \int_{\RR^+} f_D(x) \ud \tilde P_n(x)
\end{equation}
is a consistent estimator.

Furthermore, if $f$ is bounded everywhere, we can make the uninformative choice $D=\infty$.
\end{lemma}
\begin{IEEEproof} Since the tapered function $f_D$ is continuous and bounded on $\RR^+$, the almost sure weak convergence of $\tilde P_n$ to $P$ implies that $\int_{\RR^+} f_D \ud \tilde P_n \to\as \int_{\RR^+} f_D\ud P$. But since $P$ is supported on $C$ and $f_D$ agrees with $f$ on $C$, we have $\int_{\RR^+} f_D\ud P = \int_C f_D \ud P = \int_C f \ud P$.
\end{IEEEproof}

In general, however, we will be interested in problems where we do not have an \emph{a priori} knowledge about the endpoints of $C$, and where an uninformative choice cannot be made because $f$ is not bounded on $\RR^+$, such as $f(x)=\log x$, $1/x$, or $x^q$. For these problems, we can apply our methodology of integrating against $\tilde P_n$ by first establishing a rate for the convergence of equation (\ref{eq:as-weak}). We characterize such a rate using a sequence $K_n\to\infty$, such that:
\begin{equation} \label{eq:convrate}
	K_n d_\mathrm{W}(\tilde P_n,P) \to\as 0,
\end{equation}
where $d_\mathrm{W}$ denotes the Wasserstein distance, which can be expressed in its dual forms:
\begin{eqnarray}
d_\mathrm{W}(\tilde P_n,P)&\equiv& \int_{\RR^+} | F(x; P_n) - F(x; P) | dx \nonumber\\
&=& \sup_{h\in\mathrm{Lipschitz}(1)} \left| \int_{\RR^+} h \ud P_n - \int_{\RR^+} h \ud P \right|.\quad \label{eq:wassdual}
\end{eqnarray}
In the remainder of the paper we will particularly focus on $K_n$ of the form $n^s$ for some $s>0$.

In the following lemma, we describe how we can use convergence rates such as (\ref{eq:convrate}) to construct consistent estimators that work with no prior knowledge on $C$, for a large subclass of canonical problems. 

\begin{lemma} \label{lemma:setting2}
Consider a canonical problem characterized by some $f$, which is Lipschitz on every closed interval $[a,b]$, \mbox{$0<a\leq b<\infty$}. If $K_n d_\mathrm{W}(\tilde P_n,P) \to\as 0$ as $n\to \infty$, for some $K_n\to\infty$, then we can choose $D_n\to\infty$ such that
\begin{equation} \label{eq:setting2}
	\hat Y_n = \int_{\RR^+} f_{D_n}(x) \ud \tilde P_n(x)
\end{equation}
is a consistent estimator.
The growth of $D_n$ controls the growth of the Lipschitz constant of $f_{D_n}$, which should be balanced with the convergence rate $K_n$. More precisely, $\hat Y_n$ in (\ref{eq:setting2}) is consitent for any $D_n\to\infty$ that additionally satisfies
\begin{equation} \label{eq:lipratio}
	\liminf_{n\to\infty} \frac{K_n}{\mathrm{Lip}(f_{D_n})} > 0,
\end{equation}
where $\mathrm{Lip}(g)$ indicates the Lipschitz constant of $g$.
\end{lemma}
\begin{IEEEproof} First note that for any $D\geq (\check{c}^{-1}\vee\hat{c})$, since $P$ is supported on $C$ and $f_D$ agrees with $f$ on $C$, we have:
\begin{equation} \label{eq:limitequiv}
\int_{\RR^+} f_D \ud P = \int_{C} f_D \ud P = \int_{C} f \ud P.
\end{equation}

Then, using the fact that for every $D$, $f_D/\mathrm{Lip}(f_D)$ is $\mathrm{Lipschitz}(1)$, we can invoke the dual representation (\ref{eq:wassdual}) of the Wasserstein distance to write:
\begin{equation} \label{eq:wassconv}
K_n \sup_{D} \frac{1}{\mathrm{Lip}(f_D)}  \left| \int_{\RR^{+}} f_D \ud \tilde P_n - \int_{\RR^{+}} f_D \ud P\right| \to\as 0.
\end{equation} 

By combining equations (\ref{eq:limitequiv}) and (\ref{eq:wassconv}), it follows that for any sequence $D_n \to \infty$, we have:
\begin{equation} \label{eq:factorconvergence}
\frac{K_n}{\mathrm{Lip}(f_{D_n})}  \left| \int_{\RR^{+}} f_{D_n} \ud \tilde P_n - \int_{C} f \ud P\right| \to\as 0.
\end{equation}

If furthermore $D_n$ is chosen such that equation (\ref{eq:lipratio}) is satisfied, then the factor $\frac{K_n}{\mathrm{Lip}(f_{D_n})}$ is eventually bounded away from zero, and can be eliminated from equation (\ref{eq:factorconvergence}) to lead to the convergence of the estimator.
\end{IEEEproof}

Of course, there may be more than one way in which one could construct $\tilde P_n$. In this paper, we focus on demonstrating the validity and usefulness of the methodology by providing two possible constructions. The results would remain valid regardless to the specific construction, and other constructions boasting more appealing properties, such as rates of convergence under more lenient assumptions, are welcome future contributions to this framework.

\section {The Good-Turing Pseudo-Empirical Measure}
\label{sec:pseudo-empirical}

\subsection{Definitions and Properties}

The platform on which we build our estimation scheme is the Good-Turing estimator, and in particular its strong consistency established by WVK. In this section, we review the main definition and properties relevant to the rest of the development. Let $B_{n,k}$ be the subset of symbols of $A_n$ that appear exactly $k$ times in the samples $X_{n,1},\cdots,X_{n,n}$. The Good-Turing estimation problem, in reference to the pioneering work of Good in \cite{wext:good}, is the estimation of the quantities $\gamma_{n,k} = p_n(B_{n,k})$, for each $k=0,1,\cdots,n$, that is the total probability of all symbols that appear exactly $k$ times. We can group these with the notation $\gamma_n\equiv\{\gamma_{n,k}\}_{k\in\NN_0}$, which we pad with zeros for $k > n$. In particular, Good suggests the following estimator.

\begin{definition} Let $\varphi_{n,k}=|B_{n,k}|$ be the number of symbols of $A_n$ that appear $k$ times in $X_{n,1},\cdots,$ $X_{n,n}$. The \emph{Good-Turing estimator} $\phi_n\equiv\{\phi_{n,k}\}_{k\in\NN_0}$ of $\gamma_{n}$, for each $k\in\NN_0$, is
\begin{equation} \label{eq:gt-estimator}
\phi_{n,k} = \frac{(k+1) \varphi_{n,k+1}}{n}.
\end{equation}
\end{definition}

WVK establish a host of convergence properties for the Good-Turing estimation problem and the Good-Turing estimator. We group these in the following theorem. 

\begin{theorem} \label{thm:convergence}
Define the Poisson $P$-mixture  $\lambda\equiv\{\lambda_k\}_{k\in\NN_0}$ as, for each $k\in\NN_0:$
\begin{equation} \label{eq:poisson-mixture}
	\lambda_k = \int_C \frac{x^k e^{-x}}{k!} \ud P(x).
\end{equation}
We then have the following results that determine the limiting behavior of $\gamma_n$, and the strong consistency of the Good-Turing estimator $\phi_n:$
\begin{itemize}
\item[(i)]  We have that $\gamma_{n,k} \to\as \lambda_k$ and $\phi_{n,k} \to\as \lambda_k$, and therefore $|\phi_{n,k}-\gamma_{n,k}| \to\as  0$, pointwise for each $k\in\NN_0$ as $n\to\infty$.
\item[(ii)]  By Scheff\'{e}'s theorem (\cite{wext:billingsley}, p. 215), it also follows that these convergences hold in $L_1$ almost surely, in that \mbox{$\|\gamma_n-\lambda\|_1\to\as 0$} and $\|\phi_n-\lambda\|_1\to\as 0$, and therefore $\|\phi_n-\gamma_n\|_1\to\as 0$, as $n\to\infty$.
\end{itemize}
\end{theorem}

\subsection{Empirical Measure Analogy}

The analogy that we would like to make in this section is the following. Assuming $\lambda$ is given, one could take $n$ i.i.d.\ samples from it, and form the empirical measure or the type, call it $\hat \lambda_n \equiv \{\hat \lambda_{n,k}\}_{k\in\NN_0}$. Such an empirical measure would satisfy well-known statistical properties, in particular the strong law of large numbers would apply, and we would have $\hat \lambda_{n,k} \to\as \lambda_k$. By Scheff\'{e}'s theorem, $L_1$ convergence would also follow. It is evident from Theorem \ref{thm:convergence} that despite the fact that we do not have such a true empirical measure, the Good-Turing estimator $\phi_n$ behaves as one, and we may be justified to call it a \emph{pseudo-empirical measure}.

Now observe that since, for discrete distributions, the total variation distance is related to the $L_1$ distance by $\sup_{B\subset \NN_0} | \hat \lambda_n(B) - \lambda(B) | = \frac{1}{2} \| \hat \lambda_n - \lambda \|_1$, the true empirical measure also converges in total variation. As a special case, the Glivenko-Cantelli theorem applies in that \mbox{$\sup_k | F(k; \lambda) - F(k; \hat \lambda_n)| \to\as 0$}. Recall that $F(.;.)$ denotes the cumulative of the second argument (a measure) evaluated at the first argument. In light of the above, this remains valid for the pseudo-empirical measure. However, for the classical empirical measure, we also have the \emph{rate} of convergence in the Glivenko-Cantelli theorem, in the form of the Kolmogorov-Smirnov theorem and its variants for discrete distributions, see for example \cite{wext:wood-altavela}. Such results are often formulated in terms of a convergence in probability of rate $\frac{1}{\sqrt{n}}$. So we next ask whether such rates hold for the pseudo-empirical measure as well.

We first note that the rare-events source model is lenient, in the sense that it does not impose any convergence rate on $P_n \Rightarrow P$. Therefore, convergence results that aim to parallel those of a true empirical measure will depend on assumptions on the rate of this core convergence. In particular, let us assume that we know something about the weak convergence rate of $P_n$ to $P$ in terms of the Wasserstein distance, in that we assume there exists an $r>0$ such that
$$
	n^r d_\mathrm{W}(P_n,P) \to 0.
$$
For example, in Lemma \ref{lemma:quantization-rate}, we will show that this holds true for a class of rare-events sources suggested by WVK.

Next, note that Lemma 11 in WVK gives the following useful concentration rate for the pseudo-empirical measure around its mean.
\begin{lemma} \label{lemma:concentration-rate}
For any $\delta>0$, $n^{1/2-\delta} \|\phi_n-\E[\phi_n]\|_1\to\as 0$.
\end{lemma}

In the following statement, we show that a Kolmogorov-Smirnov-type convergence to $\lambda$  does hold for the pseudo-empirical measure $\phi_n$, with a rate that is essentially the slower of that of the concentration of Lemma \ref{lemma:concentration-rate} and that of the rare-events source itself.
\begin{theorem} \label{thm:k-s}
Let $r>0$ be such that $n^r d_\mathrm{W}(P_n,P) \to 0$. Then for any $\delta>0$,  we have:
\begin{equation}
n^{\min\{r,\;1/2\}-\delta} \sup_k | F(k; \lambda) - F(k; \phi_n)| \to\as 0.
\end{equation}
\end{theorem}
\begin{IEEEproof}
For convenience, define $B_k\equiv\{0,\cdots,k\}$. The proof requires three approximations. The first is to approximate $\phi_n$ with $\E[\phi_n]$. This is already achieved using Lemma \ref{lemma:concentration-rate}. Since the $L_1$ distance is twice the total variation distance, and specializing to the subsets $B_k$, we have that for all $\delta>0$:
\begin{equation} \label{eq:part-0}
n^{1/2-\delta} \sup_k | F(k; \E[\phi_n]) - F(k; \phi_n)| \to\as 0.
\end{equation}

The next two approximations are \emph{(a)} to approximate $\E[\phi_n]$ with a Poisson $P_n$-mixture (using the theory of Poisson approximation), and \emph{(b)} to approximate the latter with $\lambda$, which is a Poisson $P$-mixture (using the convergence in $d_\mathrm{W}(P_n,P)$).

\emph{Part (a) --} For convenience, let $\pi_n$ be a $\mathrm{Poisson}(x)$ $P_n$-mixture, and let $\eta_n$ be a $\mathrm{Binomial}\left(\frac{x}{n},n\right)$  $P_n$-mixture. One can show, as in the proof of Lemma 7 of WVK, that $\E[\phi_n]$ is a $\mathrm{Binomial}\left(\frac{x}{n},n-1\right)$ $P_n$-mixture. We first relate $\E[\phi_n]$ to $\eta_n$ which is the natural candidate for Poisson approximation. We then use Le Cam's theorem to relate $\eta_n$ to $\pi_n$.

We start with a general observation. Let $\mathscr{F}=\{f(\cdot;x):x\in C\}$ and $\mathscr{G}=\{g(\cdot;x):x\in C\}$ be two parametric classes of probability mass functions over $\NN_0$, e.g. Poisson and Binomial, and let $Q$ be a mixing distribution supported on $C$. Say that for some subset $B\subset \NN_0$, we have the pointwise bound \mbox{$|f(B;x)-g(B;x)|\leq \ell(x)$}. It follows that the mixture of the bound is  also a bound on the mixture. More precisely:
\begin{equation} \label{eq:observation}
	\!\!\left|\int_C f(B;x) dQ(x) - \int_C g(B;x) dQ(x) \right| \leq \int_C \ell(x) dQ(x).
\end{equation}
Note that if the pointwise bound above holds uniformly over $B$, then the same is true for the mixture bound. We will use this particularly with the subsets $B_k$, to bound the difference of cumulative distribution functions.

Now let $g_n(k;x)$ be the c.d.f.\ of a $\mathrm{Binomial}\left(\frac{x}{n},n\right)$ random variable, and let ${\tilde g}_n(k;x)$ be the c.d.f.\ of a $\mathrm{Binomial}\left(\frac{x}{n},n-1\right)$ random variable. For any given $k$, we have the following:
\begin{eqnarray*}
\lefteqn{\left(1-\frac{x}{n}\right) {\tilde g}_n(k;x)}\\
&=& \sum_{m=0}^k \frac{n-m}{n} \binom{n}{m}\left(\frac{x}{n}\right)^m \left(1-\frac{x}{n}\right)^{n-m}\\
&=& g_n(k;x) - \frac{1}{n} \sum_{m=0}^k m \binom{n}{m}\left(\frac{x}{n}\right)^m \left(1-\frac{x}{n}\right)^{n-m}.
\end{eqnarray*}

Using the facts that the sum is no larger than the mean and that ${\tilde g}_n(k;x)\leq 1$, it follows that for any given $k$ we have:
\begin{eqnarray*}
\lefteqn{\left|g_n(k;x)-{\tilde g}_n(k;x)\right|}\\
&=& \left|\frac{1}{n} \sum_{m=0}^k m \binom{n}{m} \left(\frac{x}{n}\right)^m \left(1-\frac{x}{n}\right)^{n-m} - \frac{x}{n} {\tilde g}_n(k;x)\right|\\
&\leq& \frac{x}{n}
\end{eqnarray*}

Note that $\int_C g_n(k;x) \ud P_n = F(k; \eta_n)$, the c.d.f.\ of $\eta_n$, and $\int_C {\tilde g}_n(k;x) \ud P_n = F(k; \E[\phi_n])$, the c.d.f.\ of $\E[\phi_n]$. Using the observation leading to equation (\ref{eq:observation}), it follows that:
\begin{equation} \label{eq:binomials}
\sup_k | F(k; \E[\phi_n]) - F(k; \eta_n)| 
\leq \frac{1}{n} \int_C x \ud P_n(x) \leq \frac{\hat c}{n}.
\end{equation}

Using Le Cam's theorem (see, for example, \cite{wext:steele}), we know that the total variation distance, and hence the difference of probabilities assigned to any subset $B\subset \NN_0$ by a $\mathrm{Poisson}(x)$ distribution and a $\mathrm{Binomial}\left(\frac{x}{n},n\right)$ distribution is upper-bounded by $\frac{x^2}{n}$. We apply this to the subsets $B_k$, and use the observation leading to equation (\ref{eq:observation}) once again to extend this result to the respective $P_n$-mixtures:
\begin{equation} \label{eq:lecam}
\sup_k | F(k; \pi_n) - F(k; \eta_n)| 
\leq \frac{1}{n} \int_C x^2 \ud P_n(x) \leq \frac{{\hat c}^2}{n}.
\end{equation}

By combining equations (\ref{eq:binomials}) and (\ref{eq:lecam}), we deduce that for all $\delta>0$:
\begin{equation} \label{eq:part-a}
n^{1-\delta} \sup_k | F(k; \E[\phi_n]) - F(k; \pi_n)| \to 0.
\end{equation}

\emph{Part (b) --} Now let $h(k;x)$ be the c.d.f.\ of a $\mathrm{Poisson}(x)$ random variable. Observe that:
\begin{eqnarray*}
0&\leq&\frac{\ud}{\ud x} h(k;x) = \sum_{m=0}^k -\frac{x^m e^{-x}}{m!} + m \frac{x^{m-1} e^{-x}}{m!}\\
&\leq& \frac{1}{x} \sum_{m=0}^k  m \frac{x^{m} e^{-x}}{m!} = \frac{1}{x} \E\left[\mathrm{Poisson}(x)\right] = 1.
\end{eqnarray*}

Therefore, when viewed as a function of $x$, $h(k;x)$ is a $\mathrm{Lipschitz}(1)$ function on $C$ for all $k$. Using the dual representation of the Wasserstein distance, we then have:
\begin{eqnarray*}
\lefteqn{\sup_k |F(k;\pi_n)-F(k;\lambda)|}\\
&=& \sup_k \left| \int_C h(k;x) \ud P_n(x) - \int_C h(k;x) \ud P(x) \right|\\
&\leq& \sup_{h\in\mathrm{Lipschitz}(1)} \left| \int_C h \ud P_n - \int_C h \ud P \right| = d_W(P_n,P).
\end{eqnarray*}
Using the assumption of the convergence rate of $P_n$ to $P$, it follows that for all $\delta>0$ we have:
\begin{equation} \label{eq:part-b}
n^{r-\delta} \sup_k |F(k;\pi_n)-F(k;\lambda)| \to 0.
\end{equation}
The statement of the theorem follows by combining equations (\ref{eq:part-0}), (\ref{eq:part-a}), and (\ref{eq:part-b}).
\end{IEEEproof}

In a practical situation, one would expect that the rare-events source is well-behaved enough that $r>1/2$, and that the bottleneck of Theorem \ref{thm:k-s} is given by the $1/2$ rate, and therefore we have a behavior that more closely parallels a true empirical measure. Indeed, some natural constructions obey this principle. Most trivially, for a sequence of uniform sources, e.g. if $p_n(a) = 1/n$, we have $P_n=P$, and therefore $r=\infty$. More generally, consider the following class of rare-events sources suggested by WVK.

\begin{definition} \label{def:quantization}
Let $g$ be a density on $[0,1]$ that is continuous Lebesgue almost everywhere, and  such that $\check c \leq g(w) \leq \hat c$ for all $w\in[0,1]$. Let $A_n=\{1,\cdots,\lfloor \alpha n\rfloor \}$ for some $\alpha>0$, and for every $a\in A_n$ let $p_n(a) = \int_{(a-1)/\lfloor \alpha n\rfloor}^{a/\lfloor \alpha n\rfloor} g(w) \ud w$. One can then verify that $\{(A_n,p_n)\}$ is indeed a rare-events source, with $P$ being the law of $g(W)$, where $W\sim g$. We call such a construction a \emph{rare-events source obtained by quantizing $g$}.
\end{definition}

\begin{lemma} \label{lemma:quantization-rate}
Let $g$ be a density as in Definition \ref{def:quantization}, and let $\{(A_n,p_n)\}$ be a rare-events source obtained by quantizing $g$. If $g$ has finitely many discontinuities, and is Lipschitz within each interval of continuity, then for all $r<1$:
\vspace{-2pt}
$$ n^r d_\mathrm{W}(P_n,P) \to 0$$
\end{lemma}
\begin{IEEEproof}
Without loss of generality, assume $\alpha=1$, and that the largest Lipschitz constant is $1$. Consider the quantized density on $[0,1]$:
\vspace{-8pt}
$$
g_n(w) = n \int_{(\lceil wn \rceil-1)/n}^{\lceil wn \rceil/n} g(v) \ud v,
$$
where the integral is against the Lebesgue measure. Then it follows that $P_n$ is the law of $g_n(W_n)$, where $W_n \sim g_n$.

Say $g$ has $L$ discontinuities, and let $D_n$ be the union of the $L$ intervals of the form $[(a-1)/n, a/n]$ which contain these discontinuities. In all other intervals, we have that \mbox{$|g(w)-g_n(w)|\leq 1/n$}, using Lipschitz continuity and the intermediate value theorem. It follows that
\begin{eqnarray*}
	\lefteqn{\int_{[0,1]} |g(w)-g_n(w)| \ud w}\\
&=& \int_{D_n} |g-g_n| \ud w + \int_{[0,1]\setminus D_n} |g-g_n| \ud w \leq \frac{L}{n}+\frac{1}{n}.
\end{eqnarray*}

For any particular $x\in C$, let $B_x = \{w\in [0,1]: g(w) < x\}$. We then have 
\begin{eqnarray*}
\lefteqn{| F(x;P_n) - F(x;P) | = \left| \int_{B_x} g(w)-g_n(w) \ud w \right|} \qquad \\ &\leq& \int_{B_x} |g(w)-g_n(w)| \ud w \leq \frac{L+1}{n}.
\end{eqnarray*}
By integrating over all $x$:
\begin{equation*}
d_\mathrm{W}(P_n,P) = \int_C | F(x;P_n) - F(x;P) | \ud x \leq \frac{(L+1)(\hat c - \check c)}{n}.
\end{equation*}
Therefore the lemma follows.
\end{IEEEproof}

We end by remarking that the rare-events sources covered by Lemma \ref{lemma:quantization-rate} are rather general in nature. For example, all of the illustrative and numerical examples offered by WVK are special cases (more precisely, they have piecewise-constant $g$).

\section{Constructing \texorpdfstring{$\tilde{P}_n$}{tilde Pn} via Mixing Density Estimation}
\label{sec:weakly-convergent}

We would now like to address the task of using $X_{n,1},\cdots,X_{n,n}$ to construct a sequence of probability measures $\tilde{P}_n$ that, for almost every sample sequence, converges weakly to $P$, as outlined in Section \ref{sec:solution}. Since we have established the Good-Turing estimator as a pseudo-empirical measure issued from a Poisson $P$-mixture, in both consistency and rate, this is analogous to a mixture density estimation problem, with the true empirical measure replaced with the Good-Turing estimator $\phi_n$.

We start by noting that the task is reasonable, because the mixing distribution in a Poisson mixture is identifiable from the mixture itself. This observation can be traced back to \cite{wext:teicher} and \cite{wext:feller}. Then, the first natural approach is to use non-parametric maximum likelihood estimation. In Section \ref{sec:simar}, we use Simar's work in \cite{wext:simar} to construct a valid estimator in this framework. Unfortunately, to the best of the authors' knowledge, the maximum likelihood estimator does not have a well-studied rate of convergence on the recovered mixing distribution. In Section \ref{sec:chen} we consider instead a minimum distance estimator, with which Chen gives optimal rates of convergence in \cite{wext:chen}, albeit by assuming finite support for $P$.

\subsection{Maximum Likelihood Estimator} \label{sec:simar}

We first define the maximum likelihood estimator in our setting. Despite the fact that it is not, strictly speaking, maximizing a true likelihood, we keep this terminology in light of the origin of the construction.

\begin{definition} Given the pseudo-empirical measure (Good-Turing estimator) $\phi_n$ the \emph{maximum likelihood estimator} of the mixing distribution is a probability measure $\tilde{P}_n^\textrm{ML}$ on $\RR^+$ which maximizes the pseudo-likelihood as follows:
\begin{equation}
\tilde{P}_n^\textrm{ML} \in \argmax_Q \sum_{k=0}^\infty \phi_{n,k} \log \left( \int_0^\infty \frac{x^k e^{-x}}{k!} \ud Q(x) \right).
\end{equation}
\end{definition}

It is not immediately clear whether $\tilde{P}_n^\textrm{ML}$ exists or is unique. These questions were answered in the affirmative in \cite{wext:simar}. On close examination, it is clear that these properties do not depend on whether we are using a pseudo-empirical measure instead of a true empirical measure. Hence they remain valid in our context. Next, we establish the main consistency statement.

\begin{theorem} \label{thm:MLE}
For almost every sample sequence, the sequence $\{\tilde{P}_n^\textrm{ML}\}$ converges weakly to $P$ as $n\to\infty$. We write this as $\tilde{P}_n^\textrm{ML}\Rightarrow\as P$.
\end{theorem} 
\begin{IEEEproof}
The main burden of proof is addressed by Theorem \ref{thm:convergence} in establishing the strong law of large numbers for the pseudo-empirical measure, and which is originally given in WVK's Proposition 7. Indeed, in Simar's proof (\cite{wext:simar}, Section 3.3, pp. 1203--1204), we only use the fact that $\phi_{n,k}\to\as \lambda_k$ for every $k\in\NN_0$. The rest of the proof carries over, and the current theorem follows.
\end{IEEEproof}

It is worth noting that the consistency of the maximum likelihood estimator does not even require that condition (i) in the Definition \ref{def:rare-events} of the rare-events source to hold, since Theorem \ref{thm:convergence} in fact holds without that condition. In that sense, it is very general. However, when every neighborhood of $0$ or $\infty$ has positive probability under $P$, it limits the types of functions that we can allow in the canonical problems, including sequence probabilities and entropies as discussed in WVK. When $P$ is not compactly supported, it is also difficult to establish the rates of convergence.

\subsection{Minimum Distance Estimator} \label{sec:chen}

We now define a minimum distance estimator for our setting. The reason that we suggest this alternate construction of $\tilde P_n$ is that it is useful to quantify the convergence rate to $P$, and the minimum distance estimator provides such a rate. However, it does so with the further assumption that $P$ has a finite support, whose size is bounded by a known number $m$.

Also note that the definition of the estimator circumvents questions of existence by allowing for a margin of $\epsilon$ from the infimum, and does not necessarily call for uniqueness.

\begin{definition} For a probability measure $Q$ on $\RR^+$, let $\pi(Q)$ denote the Poisson $Q$-mixture. Then, given the pseudo-empirical measure $\phi_n$, a \emph{minimum distance estimator} with precision $\epsilon$ is any probability measure ${\tilde P}_n^{\mathrm{MD},m,\epsilon}$ on $\RR^+$ that satisfies
\begin{eqnarray*}
\lefteqn{\sup_k \left| F(k; \pi({\tilde P}_n^{\mathrm{MD},m,\epsilon})) - F(k;\phi_n) \right|}\qquad\\
&\leq& \inf_Q \sup_k \left| F(k; \pi(Q)) - F(k;\phi_n) \right| + \epsilon,
\end{eqnarray*}
where the infimum is taken on probability measures supported on at most $m$ points, on $\RR^+$.
\end{definition}

We now provide the main consistency and rate results associated with such estimators.

\begin{theorem} \label{thm:min-dist}
Let $r>0$ be such that $n^{r} d_\mathrm{W}(P_n,P) \to 0$, and assume that it is known that $P$ is supported on at most $m$ points. Let ${\tilde P}_n^{\mathrm{MD},m,\epsilon_n}$ be a sequence of minimum distance estimators chosen such that $\epsilon_n  < n^{-\min\{r,1/2\}}$. Then as \mbox{$n\to\infty$}, we have that for any $\delta>0$:
\begin{equation}
 n^{\min\{r/2,1/4\}-\delta}d_\mathrm{W}\left({\tilde P}_n^{\mathrm{MD},m,\epsilon_n},P\right) \to\as 0.
\end{equation}
\end{theorem}

\noindent Remark: Since $d_\mathrm{W}$ induces the weak convergence topology, it also follows that ${\tilde P}_n^{\mathrm{MD},m,\epsilon_n} \Rightarrow\as P$.

\begin{IEEEproof}
To derive rate results in \cite{wext:chen}, Chen establishes a bound on
the Wasserstein distance between mixing distributions, using the Kolmogorov-Smirnov distance between the c.d.f.s of the resulting mixtures. For this, he first introduces a notion of strong identifiability (Definition 2, p. 225), and shows that Poisson mixtures satisfy it (Section 4, p. 228). He then shows (in Lemma 2, p. 225) that if we have strongly identifiable mixtures and if two mixing distributions have a support of at most $m$ points within a fixed compact set, such as $C$, then we can find a constant $M$ (which depends non-constructively on $m$ and $C$), such that for any two such mixing distributions $Q_1$ and $Q_2$, we have:
\begin{equation} \label{eq:wasserstein-ks}
d_\mathrm{W}\left(Q_1,Q_2\right)^2 \leq M \sup_k \left| F(k;\pi(Q_1)) - F(k;\pi(Q_2)) \right|
\end{equation}

The main burden of proof therefore falls on our Theorem \ref{thm:k-s} in establishing a Kolmogorov-Smirnov-type convergence for the pseudo-empirical measure. The argument we present next is based on Chen's proof (Theorem 2, p. 226). We have:
\begin{eqnarray*}
\lefteqn{\sup_k \left| F(k; \pi({\tilde P}_n^{\mathrm{MD},m,\epsilon_n})) - F(k;\phi_n) \right|} \qquad\\
&\leq& 
\sup_k  \left| F(k; \pi({\tilde P}_n^{\mathrm{MD},m,\epsilon_n})) - F(k;\lambda) \right|\\
&&\quad
+ \sup_k \left| F(k;\lambda) - F(k;\phi_n) \right|\\
&\leq& 2 \sup_k \left| F(k;\lambda) - F(k;\phi_n) \right| + \epsilon_n,
\end{eqnarray*}
where the final inequality is due to the definition of ${\tilde P}_n^{\mathrm{MD},m,\epsilon_n}$. By Theorem \ref{thm:k-s}, and by our choice of $\epsilon_n$, it follows that for all $\delta>0$, we have:
\begin{equation} \label{eq:decay}
n^{\min\{r,1/2\}-2\delta} \sup_k \left| F(k; \pi({\tilde P}_n^{\mathrm{MD},m,\epsilon_n})) - F(k;\phi_n) \right| \to\as 0.
\end{equation}

By combining (\ref{eq:wasserstein-ks}) and (\ref{eq:decay}), the theorem follows .
\end{IEEEproof}

Note that Chen's result can be used to show more. In particular, if we think of the true mixing distribution as residing in some neighborhood of a fixed distribution, then the convergence holds uniformly over that neighborhood. This may be interpreted as a form of robustness, but we do not dwell on it here.

\section{Applications}
\label{sec:applications}

To solve canonical problems in the setting of Lemma \ref{lemma:setting1}, when an a priori bound on $C$ is known or when $f$ is bounded on $\RR^+$,
it suffices to construct a sequence of probability measures $\tilde P_n$ that weakly converges to $P$ for almost every sample sequence. Since Theorem \ref{thm:MLE} provides such a sequence, we need not go further than that.

However, to work within the more general setting of Lemma \ref{lemma:setting2}, where no knowledge of $C$ is assumed and $f$ can be any locally Lipschitz function, we can use the result of Theorem \ref{thm:min-dist}. In this section, we start by illustrating this for some of the quantities considered by WVK. We then suggest two new applications: alphabet size and support interval estimation. We conclude by remarking on some algorithmic considerations.

\subsection{Estimating Entropies and Probabilities} \label{sec:WVKexamples}

First consider the entropy of the source $H(p_n)$, and the associated problem, in normalized form, of estimating \mbox{$Y_n^H \equiv H(p_n)-\log n$}. One can then write:
\begin{equation*}
	Y_n^H = - \int_C \log x \ud P_n(x),
\end{equation*}
and therefore, by comparing to equation (\ref{eq:canonical}) with $f_n(x)=f(x)=-log(x)$, we have a canonical estimation problem, and since $Y_n^H$ is deterministic, it is also strong. If we have a bound on $C$, we can use Lemma \ref{lemma:setting1}. Otherwise, note that on intervals of the form $[D^{-1},D]$, $\log x$ is $D$-Lipshitz. Therefore if for some $s>0$,  $n^s d_\mathrm{W}(\tilde P_n, P) \to\as 0$, as given by Theorem \ref{thm:min-dist} for example, then we can apply Lemma $\ref{lemma:setting2}$ using $D_n = n^s$. If $s$ exists but is unknown, we can still apply Lemma $\ref{lemma:setting2}$ using any sequence that is $o(n^s)$, such as $D_n = e^{\log^\epsilon n}$, for some $\epsilon>0$. The consistent estimator becomes:
\begin{equation} \label{eq:entropy}
	\hat Y_n^H \equiv - \int_{\RR^+} \log_{D_n} x \ud \tilde P_n(x).
\end{equation}

Next consider the probability of the sequence $p_n(X_{n,1},\cdots,X_{n,n})$, and the associated normalized problem of estimating $Y_n^p \equiv \frac{1}{n}\log p_n(X_{n,1},\cdots,X_{n,n})+\log n$. We have (WVK, Lemma 5):
\begin{eqnarray*}
\E[Y_n^p] &=& \E[\log p_n(X_n)]+\log n\\
&=& \int_C \log x \ud P_n(x),
\end{eqnarray*}
and therefore we also have a canonical estimation problem. Using McDiarmid's theorem, one can also show that (WVK, Lemma 6) $|\E[Y_n^p]-Y_n^p|\to\as 0$, and therefore we once again have a strong canonical estimation problem, and we can construct a consistent estimator as in the case of entropy. Referring to equation (\ref{eq:entropy}), we have $\hat Y_n^p \equiv - \hat Y_n^H$.

\subsection{Estimating the Alphabet Size}

Consider the size of the alphabet $|A_n|$. Since the model describes large, asymptotically infinite, alphabets, we look at the normalized problem of estimating $Y_n^A = |A_n|/n$. We have (cf. \cite{wext:bhat}):
\vspace{-6pt}
\begin{eqnarray*}
Y_n^A &=& \frac{1}{n} \sum_{a\in A} 1 = \sum_{a\in A} \frac{p_n(a)}{n p_n(a)} \\
&=& \int_C \frac{1}{x} \ud P_n(x).
\end{eqnarray*}
\vspace{3pt}

Once again, having a deterministic sequence of the form of (\ref{eq:canonical}) with $f_n(x)=f(x)=1/x$, it follows that $\{Y_n^A\}_{n\in\NN}$ is a strong canonical problem. If we have a bound on $C$, we can use Lemma \ref{lemma:setting1}. Otherwise, note that on intervals of the form $[D^{-1},D]$, $1/x$ is $D^2$-Lipshitz. Therefore if for some $s>0$,  $n^s d_\mathrm{W}(\tilde P_n, P) \to\as 0$, as given by Theorem \ref{thm:min-dist} for example, then we can apply Lemma $\ref{lemma:setting2}$ using $D_n = n^{s/2}$. As in Section \ref{sec:WVKexamples}, if $s$ exists but is unknown, we can still apply Lemma $\ref{lemma:setting2}$ using any sequence that is $o(n^s)$, such as $D_n = e^{\log^\epsilon n}$, for some $\epsilon>0$. The consistent estimator becomes:
\begin{equation}
	\hat Y_n^A \equiv \int_{\RR^+} x^{-1}_{D_n} \ud \tilde P_n(x).
\end{equation}

\subsection{Estimating the Support Interval}

As discussed in Section \ref{sec:examples}, estimating the support interval is not a canonical problem per se. However, we show here that we can extend the framework in a straightforward fashion to provide consistent estimators of both $\check c$ and $\hat c$.

\begin{lemma}
Let $\tilde P_n\Rightarrow\as P$ such that for some $s>0$, we have $n^s d_\mathrm{W}(\tilde P_n, P) \to\as 0$. This is particularly true under the conditions of Theorem \ref{thm:min-dist}. Given $q\neq 0$ and $D\geq 1$, let $x^q_D$ denote the $D$-tapered version of $x^q$.

If $q_n = \log n / \log\log n$ and $D_n=n^{s/(2q_n)}$, then we have: as $n\to\infty$,
\vspace{-6pt}
\begin{eqnarray*}
\left( \int_{\RR^+} x_{D_n}^{-q_n} \ud \tilde P_n(x) \right) ^{1/q_n} &\to\as& \check c\\
\mathit{and}\quad \left( \int_{\RR^+} x_{D_n}^{q_n} \ud \tilde P_n(x) \right) ^{1/q_n} &\to\as& \hat c.
\end{eqnarray*}
\end{lemma}
\begin{IEEEproof} For conciseness, let us drop the argument of the probability measures, and write $\ud P$ for $\ud P(x)$. We provide the proof only for $\check c$, since the argument is analogous for $\hat c$. Recall that $\check c$ is the essential infimum of a random variable $Z \sim P$. Therefore, for any $D\geq (\check{c}^{-1}\vee\hat{c})$, we have:
\begin{equation} \label{eq:lp-linfty-alt}
\left( \int_{\RR^+} x_D^{-q} \ud P \right) ^{1/q} \to \check c\qquad\mathrm{as}\ q\to\infty.
\end{equation} 
In the absence of a rate of convergence, we cannot simply plug in $\tilde P_n$. But since we know that $n^s d_\mathrm{W}(\tilde P_n, P) \to\as 0$, we can use the dual representation of the Wasserstein distance and the fact that for every $q$ and $D$ the function $\frac{1}{q} D^{-1-q} x_D^{-q}$ is $\mathrm{Lipschitz}(1)$ over $\RR^{+}$ to state: as $n\to\infty$,
\begin{equation} \label{eq:rate-alt}
n^s \sup_{q,D} \frac{D^{-1-q}}{q}  \left| \int_{\RR^{+}} x_D^{-q} \ud \tilde P_n - \int_{\RR^{+}} x_D^{-q} \ud P\right| \to\as 0.
\end{equation} 

We now want to relate this to the difference of the $q^\mathrm{th}$ roots. Note that each of the integrals in (\ref{eq:rate-alt}) is bounded from below by $D^{-q}$. Using this and the fact that for any $a$ and $b>0$ we have $\left| a^{1/q} - b^{1/q} \right| \leq \frac{1}{q} (a\wedge b)^{\frac{1}{q}-1} \left|a-b\right|$, we can write:
\begin{eqnarray*} 
\lefteqn{\left| \left(\int_{\RR^+} x_D^{-q} \ud \tilde P_n\right)^{1/q} - \left(\int_{\RR^+} x_D^{-q} \ud P\right)^{1/q} \right|} \; \nonumber \\
&\leq& D^{2q} \cdot \ 
\frac{D^{-1-q}}{q}
\left| \int_{\RR^+} x_D^{-q} \ud \tilde P_n - \int_{\RR^+} x_D^{-q} \ud P\right|.
\end{eqnarray*}

The choices $q_n = \log n / \log\log n$ and $D_n =n^{s/(2q_n)}$, allow us to have $D_n^{2q_n} = n^s$, and yet guarantee that as $n\to\infty$ both $q_n$ and $D_n\to\infty$. With this, we can use the convergence of equation (\ref{eq:rate-alt}), to state: as $n\to\infty$,
\begin{eqnarray} \label{eq:convergence-alt}
\!\!\!\left| \left(\int_{\RR^+} x_{D_n}^{-q_n} \ud \tilde P_n\right)^{1/q_n} - \left(\int_{\RR^+} x_{D_n}^{-q_n} \ud P\right)^{1/q_n} \right| \qquad\quad\quad&& \nonumber\\
\leq n^s \frac{D_n^{-1-q_n}}{q_n}
\left| \int_{\RR^+} x_{D_n}^{-q_n} \ud \tilde P_n - \int_{\RR^+} x_{D_n}^{-q_n} \ud P\right|\to\as 0.&&
\end{eqnarray}

We then combine (\ref{eq:lp-linfty-alt}) and (\ref{eq:convergence-alt}) to complete the proof. 
\end{IEEEproof}

\begin{remarks*} Note the following:
\begin{itemize}
\item[(i)] Other scaling schemes can be devised for $q_n$ and $D_n$, as long as they both grow to $\infty$ as $n\to\infty$, yet $D_n^{2q_n}$ remains at most $\mathcal{O}\left(n^s\right)$.
\item[(ii)] If a bound $[D_{\min},D_{\max}]\supset C$ is already known, then we can taper $x^q$ accordingly, without growing $D_n$. In this case, we can also speed up the rate of convergence by choosing $q_n= \frac{s}{2} \log n/\log \frac{D_{\max}}{D_{\min}}$.
\item[(iii)] If only an upper bound or only a lower bound is known, we can taper $x^q$ accordingly, and only grow/shrink the missing bound. In this case we leave $q_n=\log n/\log \log n$ as in the Lemma.
\item[(iv)] In the Lemma and the alternatives in these remarks, if $s$ is unknown we can replace it wherever it appears (together with constant factors) with a suitably decaying term, that guarantees the behavior of remark (i). For example, in the Lemma, we can choose $D_n = n^{1/(q_n \sqrt{\log\log n})}$, since then  $D_n^{2q_n}$ becomes $o(n^s)$ for any $s$, and the proof applies. 
\end{itemize}
\end{remarks*}

\subsection{Algorithmic Considerations}

One of the appealing properties of the maximum likelihood estimator is that, by a result of Simar in \cite{wext:simar}, it is supported on finitely many points. Simar also suggests a particular algorithm for obtaining the $\tilde P_n^\mathrm{MLE}$, the convergence of which was later established in \cite{wext:bohning}, with further improvements. One can also solve for the MLE using the EM algorithm, as reviewed in \cite{wext:redner-walker}. Penalized variants are also suggested, such as in \cite{wext:leroux}. The literature on the non-parametric maximum likelihood estimator for mixtures is indeed very rich. As for the minimum distance estimator, in \cite{wext:chen} Chen suggests variants of the work in \cite{wext:deely-kruse}, where they use algorithms based on linear programming.


\bibliographystyle{IEEEtran}
\bibliography{wextbib1}

\end{document}